\documentclass[epj]{svjour}
\usepackage{amsmath}
\usepackage{graphicx}
\begin{document}
\title{Photon asymmetry measurements of $\boldmath{\overrightarrow{\gamma} \mathrm{p} \rightarrow \pi^{0} \mathrm{p}}$\\ for 
E$_{\gamma}$=320$-$650 MeV.}

\author{S. Gardner\inst{1}
\thanks{\emph{Corresponding Author:}  s.gardner.1@research.gla.ac.uk}
\and D. Howdle\inst{1}
\and M.H. Sikora\inst{6}
\and Y. Wunderlich\inst{4}
\and S. Abt\inst{10}
\and P. Achenbach\inst{2}
\and F. Afzal\inst{4}
\and P. Aguar-Bartolome\inst{2}
\and Z. Ahmed\inst{14}
\and J.R.M. Annand\inst{1}
% ...
(The A2 Collaboration at MAMI)
}                     
%
%
\institute{SUPA, School of Physics and Astronomy, University of Glasgow, Glasgow G12 8QQ, UK
\and Institut f\"ur Kernphysik, University of Mainz, D-55099 Mainz, Germany
\and Kent State University, Kent, Ohio 44242, USA 
\and Helmholtz-Institut f¨ur Strahlen- und Kernphysik, University of Bonn, D-53115 Bonn, Germany
\and INFN Sezione di Pavia, I-27100 Pavia, Italy 
\and SUPA, School of Physics, University of Edinburgh, Edinburgh EH9 3JZ, UK 
\and The George Washington University, Washington, DC 20052, USA
\and University of California Los Angeles, Los Angeles, California 90095-1547, USA 
\and Lebedev Physical Institute, 119991 Moscow, Russia
\and Institut f\"ur Physik, University of Basel, CH-4056 Basel, Switzerland 
\and Mount Allison University, Sackville, New Brunswick E4L3B5, Canada 
\and Institute for Nuclear Research, 125047 Moscow, Russia 
\and Rudjer Boskovic Institute, HR-10000 Zagreb, Croatia 
\and University of Regina, Regina, SK S4S 0A2 Canada
\and Joint Institute for Nuclear Research,141980 Dubna, Russia
\and Department of Astronomy and Physics, Saint Marys University, Halifax, Canada
\and University of Massachusetts Amherst, Amherst, Massachusetts 01003, USA
}                     
\date{Received: date / Revised version: date}
%
\abstract{
High statistics measurements of the photon asymmetry $\mathrm{\Sigma}$ for the $\overrightarrow{\gamma}$p$\rightarrow\pi^{0}$p reaction have been made in the center of mass energy range W=1214-1450 MeV. The data were measured with the MAMI A2 real photon beam and Crystal Ball/TAPS detector systems in Mainz, Germany. The results significantly improve the existing world data and are shown to be in good agreement with previous measurements, and with the MAID, SAID, and Bonn-Gatchina predictions. We have also combined the photon asymmetry results with recent cross-section measurements from Mainz to calculate the profile functions, $\check{\mathrm{\Sigma}}$ (= $\sigma_{0}\mathrm{\Sigma}$), and perform a moment analysis. Comparison with calculations from the Bonn-Gatchina model shows that the precision of the data is good enough to further constrain the higher partial waves, and there is an indication of interference between the very small $F$-waves and the $N(1520) 3/2^{-}$ and $N(1535) 1/2^{-}$ resonances.
} 
\maketitle


\section{Introduction}
\label{introduction}
\setcounter{equation}{0}
Quantum Chromodynamics (QCD) successfully describes many of the phenomena associated with elementary particles at high energies. Yet, our understanding of the low-energy, non-perturbative regime is more limited, demonstrated by our lack of precise knowledge of the excitation spectra of nucleons and mesons. The properties of baryon resonances have been mainly determined from the results of pion-nucleon scattering analyses \cite{Olive-1674-1137-38-9-090001}, with other reactions helping to fix branching ratios and photo-couplings. Beyond elastic pion-nucleon scattering, single-pion photo-production remains the most studied source of resonance information \cite{adlarson2015measurement} and many recent efforts have been directed towards obtaining complete, or nearly complete, measurements in meson-nucleon photo-production reactions using double-polarization observables \cite{Sikora-PhysRevLett.112.022501,Hornidge-PhysRevLett.111.062004}. However, high statistics measurements of single-polarization observables, over a wide photon energy and angular range remain vitally important in determining the photoproduction amplitudes from which the underlying resonance information may be extracted.

This work exploits the linearly polarized, tagged photon beam at the MAMI 1.6 GeV electron microtron in Mainz to provide beam asymmetry measurements for beam energies, $E_{\gamma}=$ 0.32 - 0.65 GeV, corresponding to a center-of-mass energy range of $W$=1.214-1.450 GeV. As shown below, the results are is in good agreement with previous measurements from Mainz, GRAAL, and Yerevan in regions where there is overlap, and provides new high statistics measurements in kinematic regions not covered by these previous experiments. In addition the new results have much finer binning in $W$ (typically 3 MeV). They are compared with the results of the partial wave analysis (PWA) fits from MAID, SAID, and Bonn-Gatchina models \cite{kamalov2008recent,Arndt2003said,Anisovich2010photoproduction}.

A discussion of the experimental arrangements for these measurements is presented in sect. \ref{experiment}. An overview of the methods used to extract the photon beam asymmetry is given in sect. \ref{analysis}. This is followed in sect. \ref{Results} by a discussion of the results of the beam asymmetry, $\mathrm{\Sigma}$, measurements. The conclusions drawn from this work are presented in sect. \ref{conclusion}.


\section{Experiment}
\label{experiment}

The photon asymmetry $\mathrm{\Sigma}$ for the reaction $\overrightarrow{\gamma}$p$\longrightarrow\pi$$^{0}$p
was measured using the Crystal Ball (CB) \cite{Starostin-PhysRevC.64.055205} as a central calorimeter and TAPS \cite{Gabler1994} as a forward calorimeter. These detectors were installed at the energy-tagged photon beam produced by bremsstrahlung from the electron beam of the 1.6 GeV Mainz Microtron (MAMI) \cite{kaiser20081,Jankowiak2006}.

The CB detector is a sphere consisting of 672 optically isolated NaI(Tl) crystals, shaped as truncated triangular pyramids, which point toward the center of the sphere. The crystals are arranged in two hemispheres that cover 93\% of 4$\pi$ sr, sitting outside a central spherical cavity with a radius of 25 cm, in which the target and inner detectors are located. In the present experiment, the 384 hexagonal cross section BaF$_{2}$ crystals of TAPS were arranged as a forward detector wall. It was installed 1.5 m downstream of the CB center and covered the full azimuthal range for laboratory polar angles from 1$^{\circ}$ to 20$^{\circ}$. More details on the calorimeters and their resolutions are given in ref. \cite{2004crystal} and references therein.

The present measurements were made in October 2008 and used a 1508 MeV electron beam from the Mainz Microtron, MAMI-C \cite{kaiser20081,Jankowiak2006}.  The energies of the incident photons were analyzed by detecting post-bremsstrahlung electrons in the Glasgow-Mainz tagged-photon spectrometer \cite{ANTHONY1991230,HALL1996698,mcgeorge2008upgrade}. The photon beam was incident on a 5 cm long liquid hydrogen (LH$_{2}$) target located in the center of the CB. The uncertainty in the energy of the tagged photons was mainly given by the width of the tagger focal-plane detectors in combination with the energy of the MAMI electron beam used in experiments. For the MAMI energy of 1508 MeV such an uncertainty was typically $\pm$ 2 MeV. The systematic uncertainty in the absolute value of E$_{\gamma}$, which is dominated by the energy calibration of the tagger, was about 0.5 MeV \cite{mcgeorge2008upgrade}.

The linear polarization of the photons was produced from coherent bremsstrahlung \cite{Timm1969coherent,lohmann1994linearly}, where the electron beam scatters coherently from a suitably aligned crystal radiator.  A thin diamond crystal (30 $\mu m$), with low mosaic-structure, was used to minimize the energy smearing of the coherent spectrum arising from electron multiple scattering effects and crystal defects in the lattice \cite{kellie2005selection}. The alignment of the diamond was carried out using the Stonehenge technique \cite{Livingston2006stonehenge} and the two orthogonal plane orientations were chosen to be at azimuthal angles of $\pm$45$^{\circ}$ with respect to the equatorial plane of the CB detector. A 2 mm diameter Pb collimator was installed 2.5 mm downstream of the radiator to enhance the ratio of coherent to incoherently scattered photons that reached the target, and to increase the degree of linear polarization. The photon polarization ranged from 4\% at E$_{\gamma}=$ 320 MeV to a maximum of 53\% at 632 MeV. 

\begin{figure}
\includegraphics[width=0.50\textwidth]{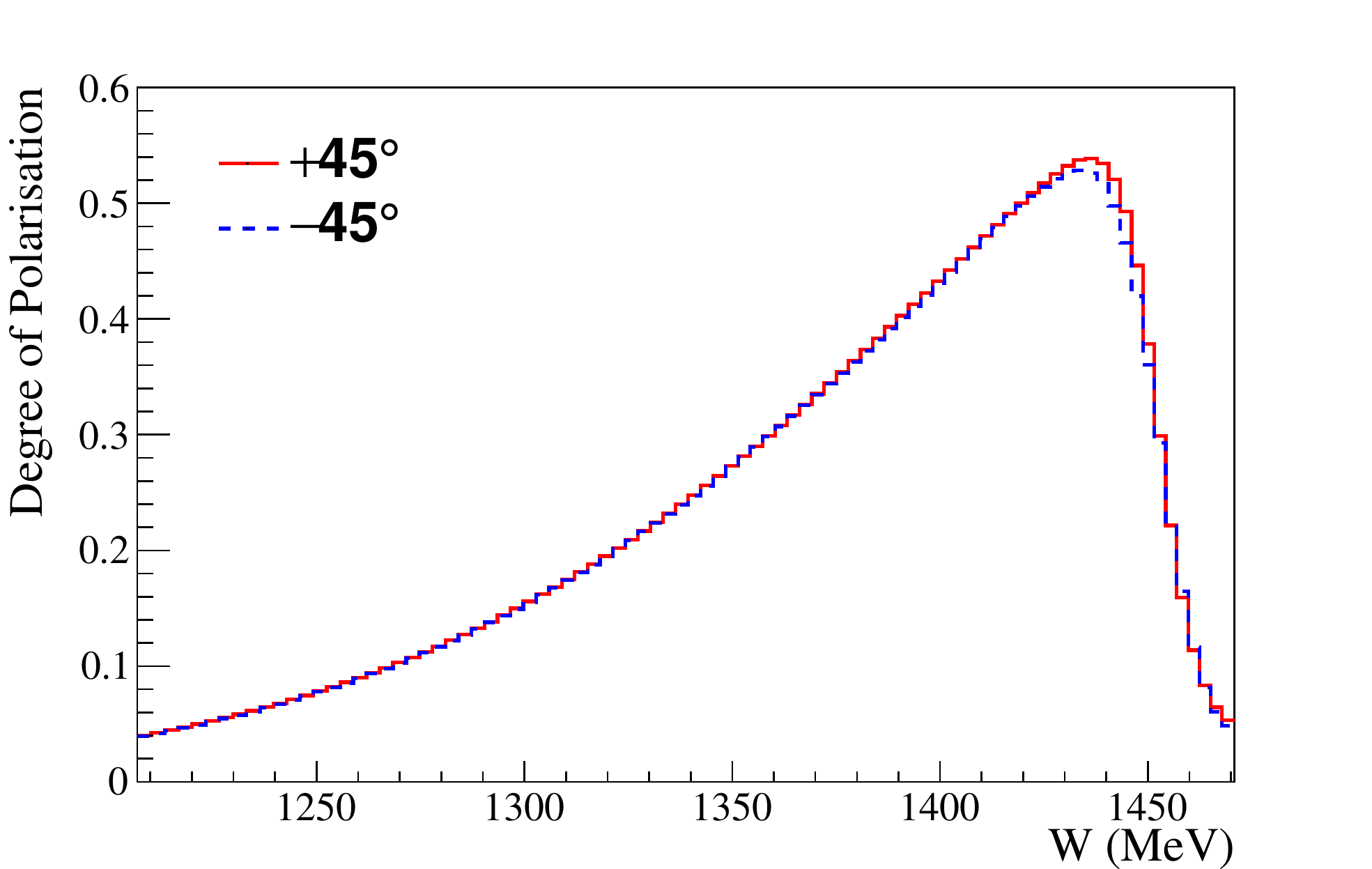}
\caption{(Color online) Degree of linear photon polarization, for the two diamond crystal orientations shown over the range of the coherent peak in center-of-mass energy $W$.}
\label{fig:polarisation}
\end{figure}

The degree of polarization was determined through constructing the \textit{enhancement} (the ratio of the E$_{\gamma}$ spectrum from the diamond radiator to that from an amorphous radiator) and fitting the enhancement with a coherent bremsstrahlung calculation. This technique gives a reliable shape for the polarization spectrum as a function of $W$ (see Fig. \ref{fig:polarisation}), but was found to have a systematic uncertainty of 5-10\% in the overall scaling due to the uncertainty in the baseline. To improve on this, we used the well-determined value of the photon beam asymmetry $\mathrm{\Sigma}$ in the region of the $\mathrm{\Delta}(1232)$ (1225 $< W <$ 1278) and limited the $\pi$$^{0}$ angular distribution to (0.2 $< \cos\theta_{cm} <$ 0.6) where the predictions from all the PWA models are in excellent agreement. Figure \ref{fig:polarisation} shows the final polarization obtained for both crystal orientations. The systematic uncertainty in the polarization was estimated to be 2\% of the magnitude of the polarization (i.e., $\mathrm{\Delta}P = \pm 0.02 P$, where $P$ is the degree of photon polarization). This was calculated using a comparison over all measured photon energies and $\pi$$^{0}$ angles with the PWA solutions thereby determining the systematic uncertainty, including uncertainties in the normalization and shape of the polarization peak.

\section{Data Analysis}
\label{analysis}

The photon asymmetry $\mathrm{\Sigma}$ has been determined as a function of 
$W$ and $\cos\theta_{cm}$, where $\theta_{cm}$ is the polar angle of the $\pi^{0}$ produced in the center-of-mass frame of the proton and the incident photon. The reaction channel was identified by detecting a single $\pi^{0}$ in coincidence with a tagged photon and selecting events with a missing mass consistent with the proton. In terms of Lorentz vectors the missing vector, $\boldmath P_{\textrm{miss}}$ is 
\begin{equation}
\boldmath P_{\textrm{miss}} =  P_{p} +  P_{\gamma} -  P_{\pi^0}\mathrm{,}
\label{eqn:mmiss}
\end{equation}
where {\boldmath $P_{p}\mathrm{,} P_{\gamma}\mathrm{,} P_{\pi^0}$} are the 4-vectors of the target proton, tagged photon and detected $\pi^{0}$, respectively.

The identification of the $\pi^{0}$ was performed by reconstruction from its 2$\gamma$ decay products. The CB and TAPS arrangement provides a very high angular acceptance for the $\pi^{0}$ decay photons that create electromagnetic showers in the crystals. This in turn leads to a high detection efficiency over most of the $\pi^{0}$ production phase space. 

Events were selected if they had two or three signal clusters in neighboring detector crystals. These were all initially assumed to be photons. Candidate $\pi^{0}$ mesons were tested by iterating over the various combinations of clusters. The $\pi^{0}$ 4-vector was constructed from the pair with an invariant mass closest to the pion mass, and the 3$^{\rm{rd}}$ cluster, if present, was assumed to be from the recoiling proton and not used further in the analysis. Incorrect combinations that may arise are cut, or subtracted, in the sWeights analysis \cite{pivk2005statistical} of the missing mass spectra outlined below. The reason for not utilizing the proton cluster was so as not to be limited by its acceptance, particularly for events with low proton momenta where the proton does not leave the target cell. 

The missing mass was constructed taking the mass of the missing 4-vector described in eq. \ref{eqn:mmiss} and a signal-background separation was performed using the $_s\mathcal{P}lot$ technique \cite{pivk2005statistical}. This is a statistical tool used to disentangle contributions from different species of events (e.g. signal and background) from observable distributions. 

The $_s\mathcal{P}lot$ analysis first fitted (unbinned extended maximum likelihood method) discriminatory variables with appropriate probability distribution functions (PDFs) to allow the determination of the yields of different species of events as a function of the discriminatory variables. The sWeights were then calculated from the covariance matrix of the fit. These steps were performed using the $_s\mathcal{P}lot$ class in the CERN ROOT RooStats package \cite{rooStats2011}. 

Specifically, the discriminatory variables used were: the coincidence time between the photon tagger and the detected $\pi^{0}$ decay photons in the CB and TAPS; and {\boldmath $P_{\textrm{miss}}$} (Eq. \ref{eqn:mmiss}).

\begin{figure}
\includegraphics[width=0.50\textwidth]{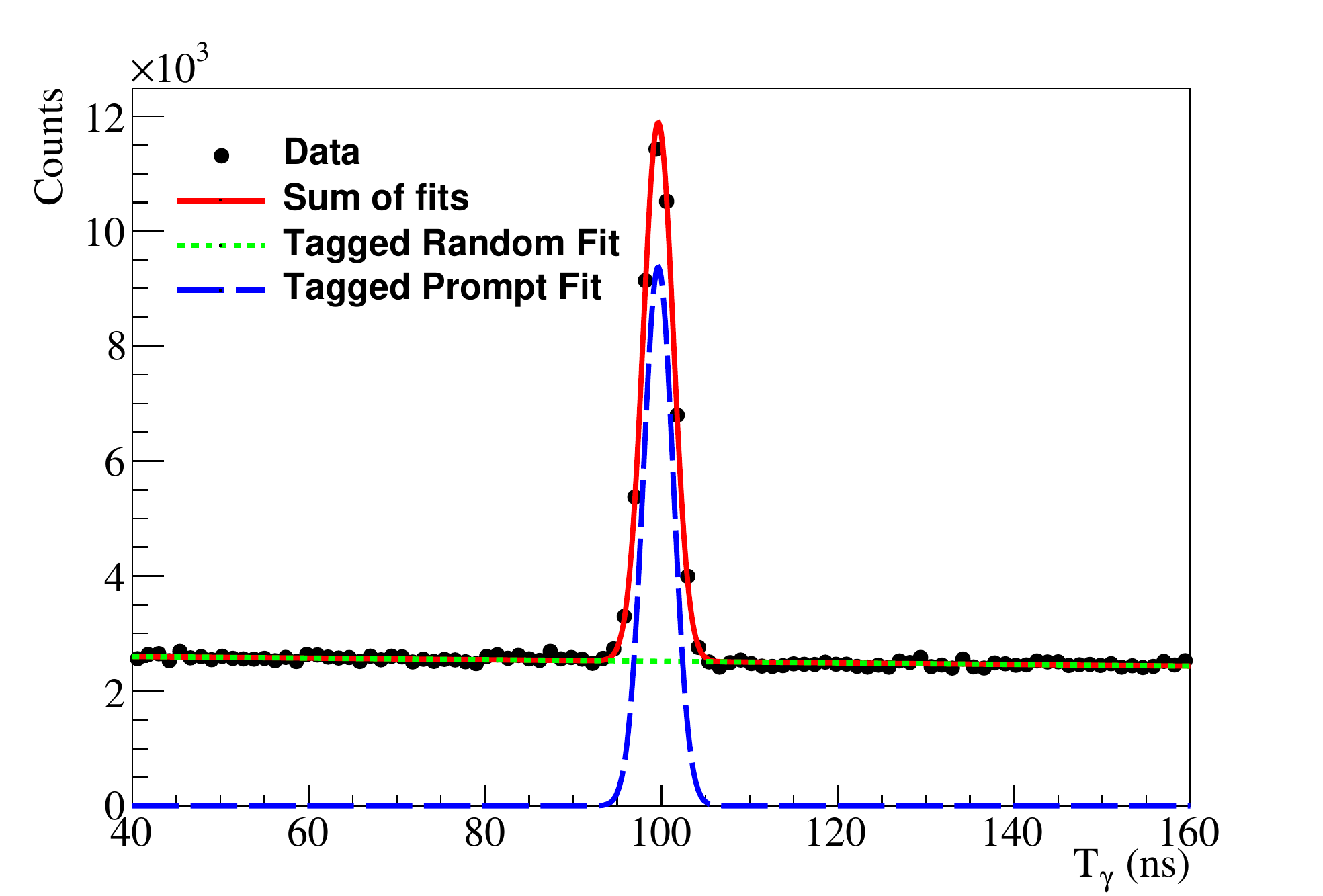}
\caption{(Color online) Timing fits used to calculate the sWeights for random subtraction. The events are divided into prompt (Blue Gaussian PDF) and random (Green Line PDF) with the sum of these PDFs (red) fitting the experimental data (black). The example illustrated is a fit for a single photon energy - pion polar angle bin $\left(W=1327\textrm{MeV},\theta_{cm}=90^{\circ}\right)$.}
\label{fig:splot-time}
\end{figure}

\begin{figure}
\includegraphics[width=0.50\textwidth]{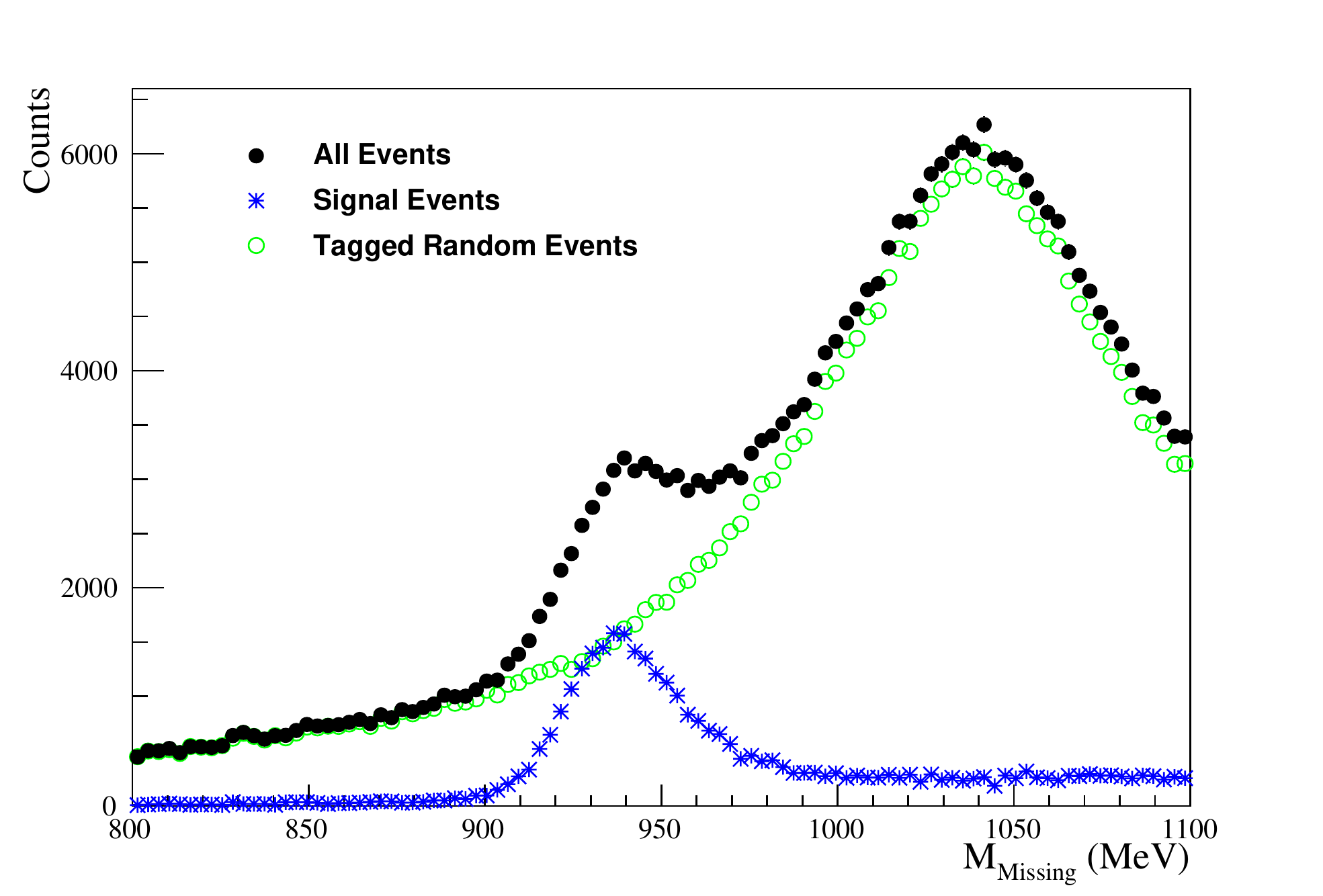}
\caption{(Color online) Missing mass from events weighted by the sWeights calculated from the timing fit. The weighted events are divided into prompt (blue) and random (green) with the total data (black). The example illustrated is a fit for a single photon energy - pion polar angle bin $\left(W=1327\textrm{MeV},\theta_{cm}=90^{\circ}\right)$.}
\label{fig:splot-timeMM}
\end{figure}

Two consecutive $_s\mathcal{P}lot$ fits are carried out on the data. Initially the coincidence time is fitted with a simple model of a Gaussian signal on a linear background. Figure \ref{fig:splot-time} shows the initial fit selecting tagged photons which are prompt to the trigger. Prompt timing sWeights calculated from the $_s\mathcal{P}lot$ are applied to the data, reproducing the missing mass spectrum found in fig. \ref{fig:splot-timeMM}.

The second fit is made to the missing mass distribution using PDFs from Geant4\cite{Agostinelli2003geant4} simulations of contributing reaction channels. Considered background reaction channels include Compton scattering and photoproduction of two pion combinations off the proton. The PDFs are given some freedom to fit the experimental data. An additional PDF is constructed from data collected during experimental runs with an empty target to account for events originating from the target cell.

Parameters for adjusting the simulated PDF shapes and yields were left as free parameters in the first iteration with the exception of the empty target which is given a constant scale yield related to the relative total flux in empty target and production runs. Fits were performed for every E$_{\gamma}$ and $\cos{\theta_{cm}}$ bin for which the photon asymmetry is determined. A further fit is conducted where background PDFs have been summed together and only signal and background yields are left as free parameters. The resulting covariance matrix and values of the PDFs for a given event were used to calculate the sWeight for each event as described in \cite{pivk2005statistical}. 

\begin{figure}
\includegraphics[width=0.50\textwidth]{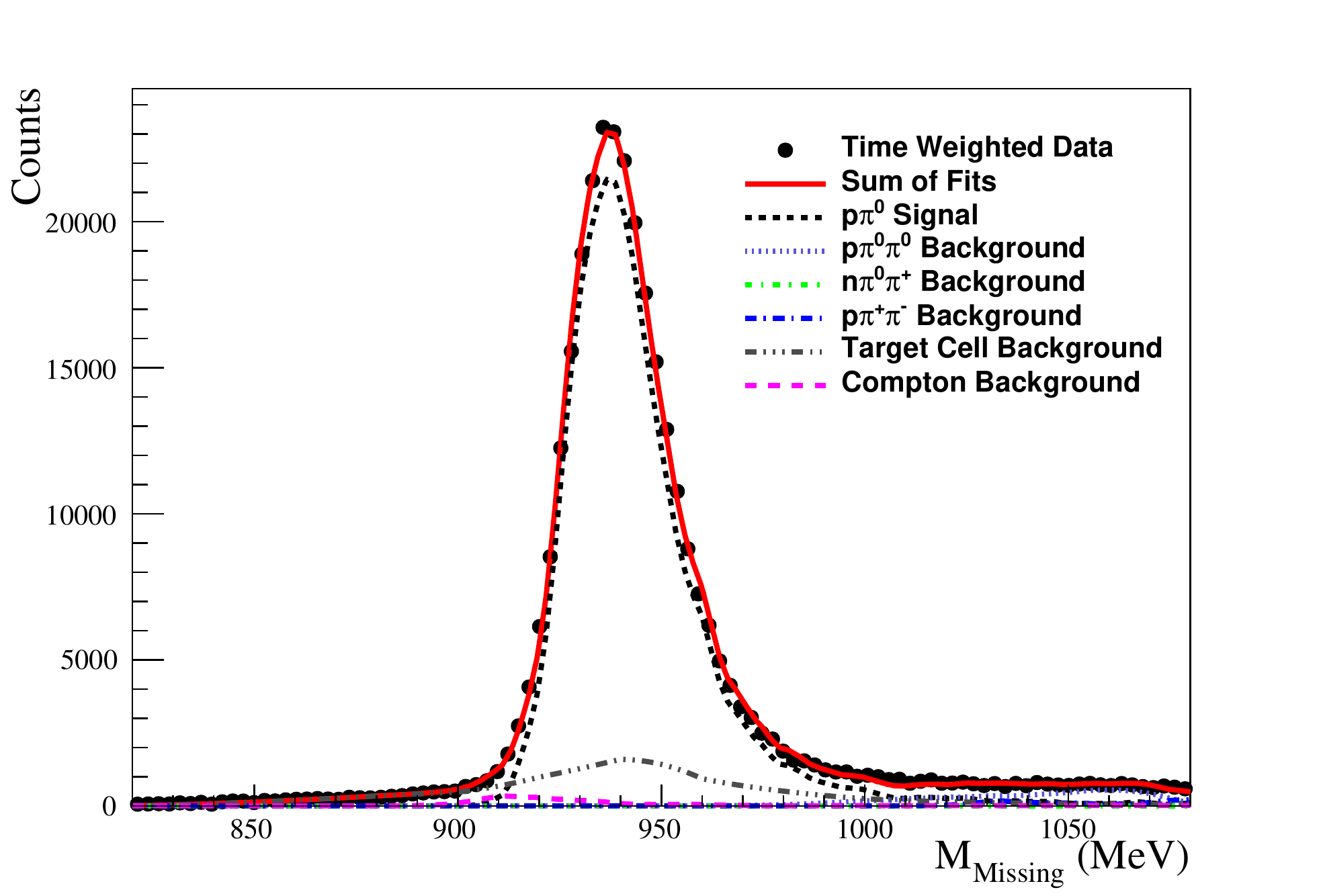}
\caption{(Color online) Projections of fits used to calculate the weights for $_s\mathcal{P}lot$. The events are divided into signal, background, and random events. The sum of these PDFs fits the experimental data very well. The example illustrated is a fit for a single center of mass energy - pion polar angle bin $\left(W=1246\textrm{MeV},\theta_{cm}=90^\circ\right)$.}
\label{fig:splot-proj}
\end{figure}

\begin{figure}
\includegraphics[width=0.50\textwidth]{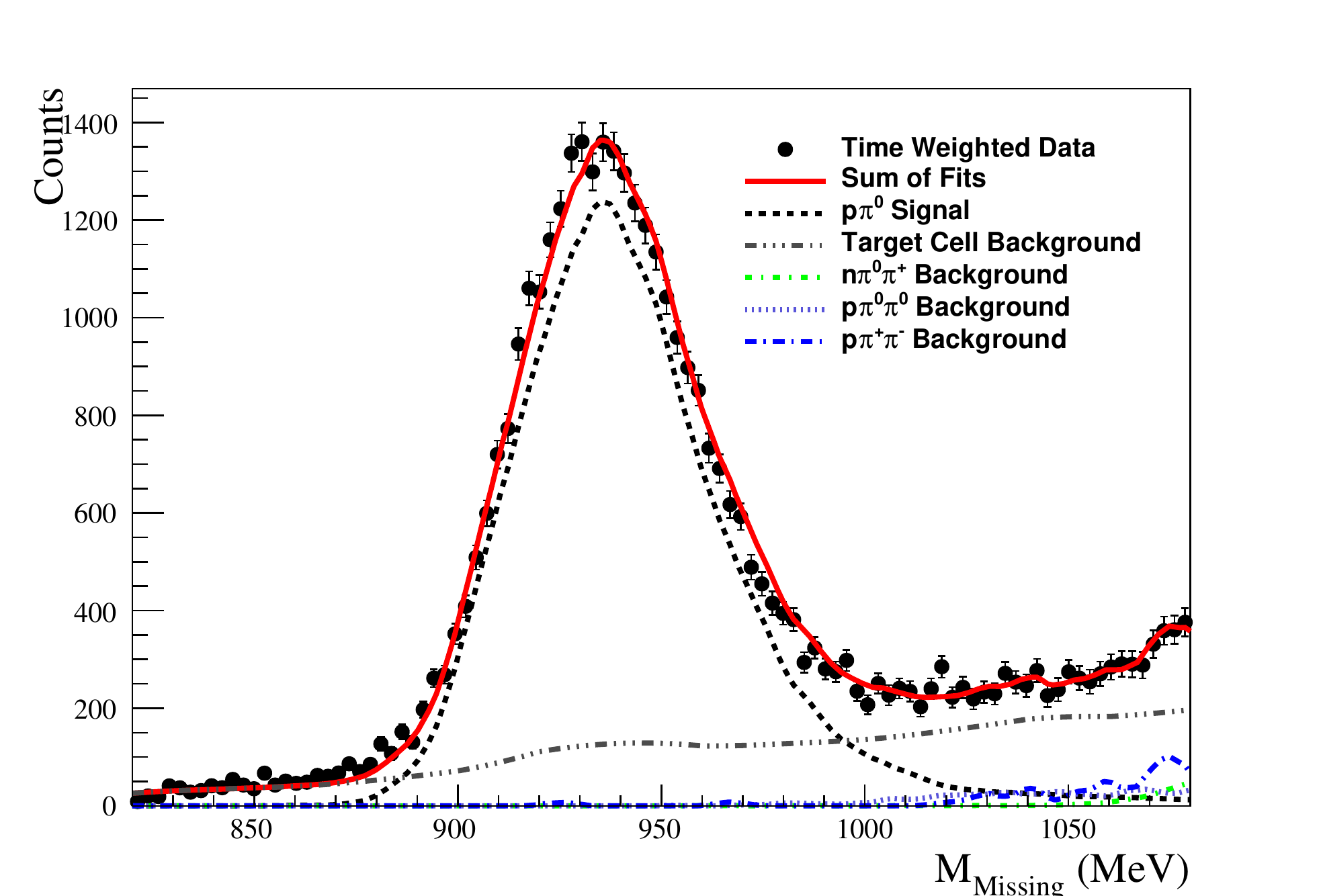}
\caption{(Color online) Projections of fits used to calculate the weights for $_s\mathcal{P}lot$. The events are divided into signal, background, and random events. The sum of these PDFs fits the experimental data very well. The example illustrated is a fit for a single center of mass energy - pion polar angle bin $\left(W=1421\textrm{MeV},\theta_{cm}=114^\circ\right)$.}
\label{fig:splot-proj2}
\end{figure}

Examples showing the results of the fits to the yields are displayed in Figs. \ref{fig:splot-proj} and \ref{fig:splot-proj2}. These figures show how the missing-mass distribution is split into the different event contributions.

\begin{figure}
\includegraphics[width=0.49\textwidth]{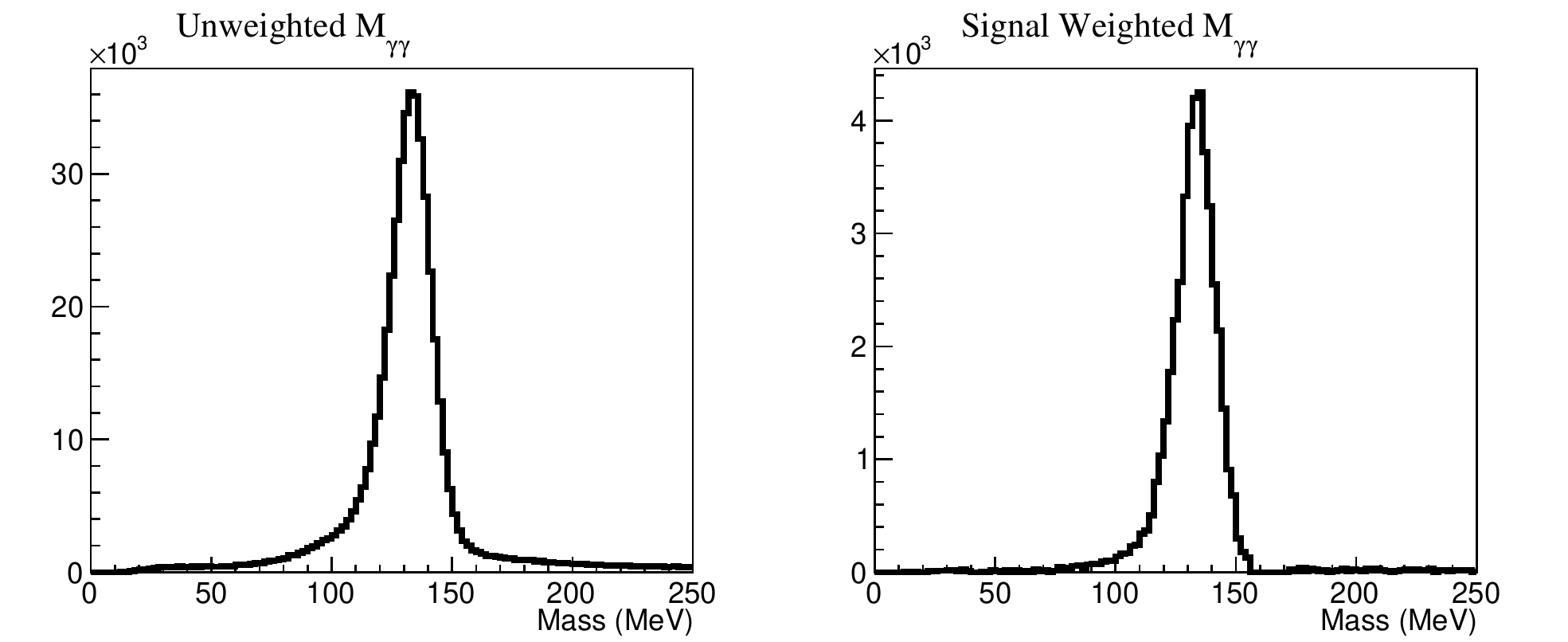}
\caption{2$\gamma$ invariant-mass distributions for all events, unweighted (left) and weighted with the $_s\mathcal{P}lot$ signal weights (right).}
\label{fig:inv-mass}
\end{figure}

Using the weights derived from this $_s\mathcal{P}lot$ fit, signal events are separated from all sources of background. For example, fig. \ref{fig:inv-mass} shows a histogram of the 2$\gamma$ invariant-mass distribution for all events and events weighted with the calculated signal. The result is a clean peak at the expected $\pi^{0}$ mass, compared with a peak on a background distribution. A low mass tail remains in the signal weighted events due to calorimeter shower loss. For analysis of the photon asymmetry, these weights were used to produce signal distributions for the $\pi^{0}$ production azimuthal angle.

The photon asymmetries $\mathrm{\Sigma}$ were extracted using the equation

\begin{equation}
\mathrm{\frac{Y_{+}(\phi)-Y_{-}(\phi)}{P_{+}Y_{+}(\phi)+P_{-}Y_{-}(\phi)} = A + \Sigma cos(2(\phi-\phi_0)),}
\label{eq:sigmaextraction}
\end{equation}
where $\mathrm{P_{+,-}}$ are the degrees of photon linear polarization scaling each event by values averaged over the full beamtime; $\mathrm{Y_{+,-}(\phi)}$ are the normalized azimuthal distributions taken with two orthogonal photon polarization orientations at $\pm45^{\circ}$ to the laboratory horizontal plane; $\phi$ is the azimuthal angle of the the plane containing the pion and recoil proton momenta, defined anti-clockwise around the beam direction from horizontal. The phase constant $\phi_{0}$ aligns the polarization of $\mathrm{Y_{+}(\phi)}$ parallel to $\phi=0$ in the detector coordinate system and A provides for the possibility of a small systematic uncertainty in the normalization of the yields.

A small dilution of the asymmetry within the signal events was expected due to the detector resolution and analysis procedure. This dilution was measured by simulating $\pi^{0}$ events with a $\mathrm{\Sigma}$ of 1 with perpendicular polarization across the kinematic range. The resulting asymmetry was measured between 0.97 and 1 across $\cos\theta_{cm}$ with no significant variation with $W$. The dilution was divided out of the final asymmetry.
    
For $\mathrm{\Sigma}$ measurements, systematic uncertainties in the location of detector systems, detector efficiencies, target density, \textit{etc.} cancel. The uncertainty arising from the flexibility in the simulation PDFs was estimated by performing repeated fits and fits with varied limits on the accepted range of missing mass. The resulting $\mathrm{\Sigma}$ values were found to be consistent within 3\%.  In addition the photon polarization P was the other significant source of systematic uncertainty, which in this case resulted in $\pm2\%$ in $\mathrm{\Sigma}$ as discussed in sect. \ref{experiment}.


\section{Results and Interpretation}
\label{Results}

\begin{figure*}
\includegraphics[width=1\textwidth]{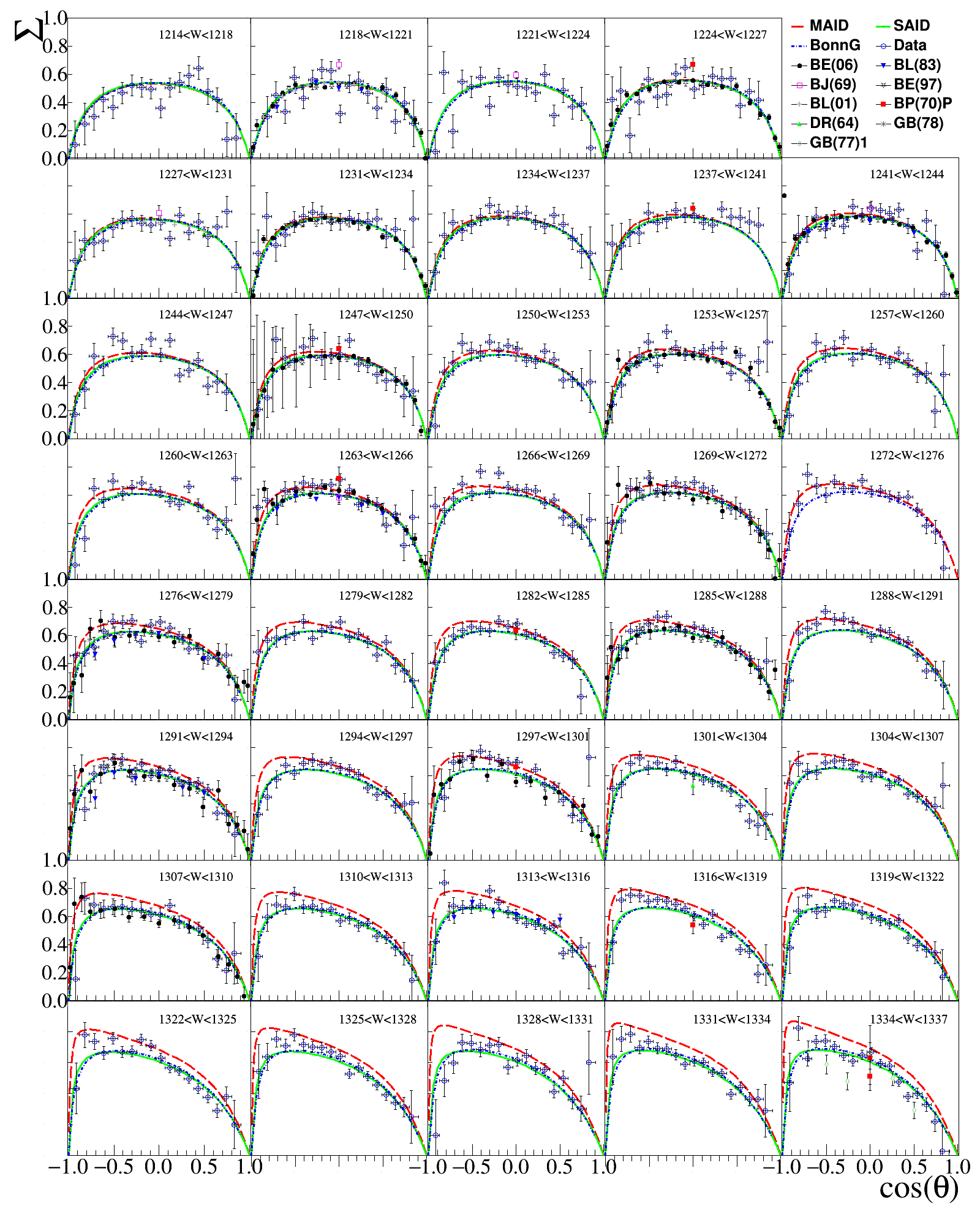}
\caption{(Color online) Photon asymmetry (open blue circles) as a function of $\cos\theta_{cm}$. The $W$ range for each plot is shown on the top right. Predictions from PWAs (MAID \cite{kamalov2008recent}, SAID \cite{DCSA2}, Bonn-Gatchina \cite{Anisovich2010photoproduction}) are shown as colored lines and results from previous experiments (BE(97) \cite{Beck1997measurement}, BE(06) \cite{Beck2006experiments}, BJ(69) \cite{Barbiellini-PhysRev.184.1402}, BL(83) \cite{Belyaev1983experimental}, BL(92) \cite{blanpied1992p}, BL(01) \cite{blanpied2001editorial}, BP(70)P \cite{zdarko1972asymmetry}, DR(64) \cite{drickey1964neutral}, GB(78) \cite{gorbenko1978measurement}, GB(77)1 \cite{gorbenko1977double}) see legend, are taken from the SAID database \cite{saidweb}.}
\label{fig:SigmaW1}
\end{figure*}

\begin{figure*}
\includegraphics[width=1\textwidth]{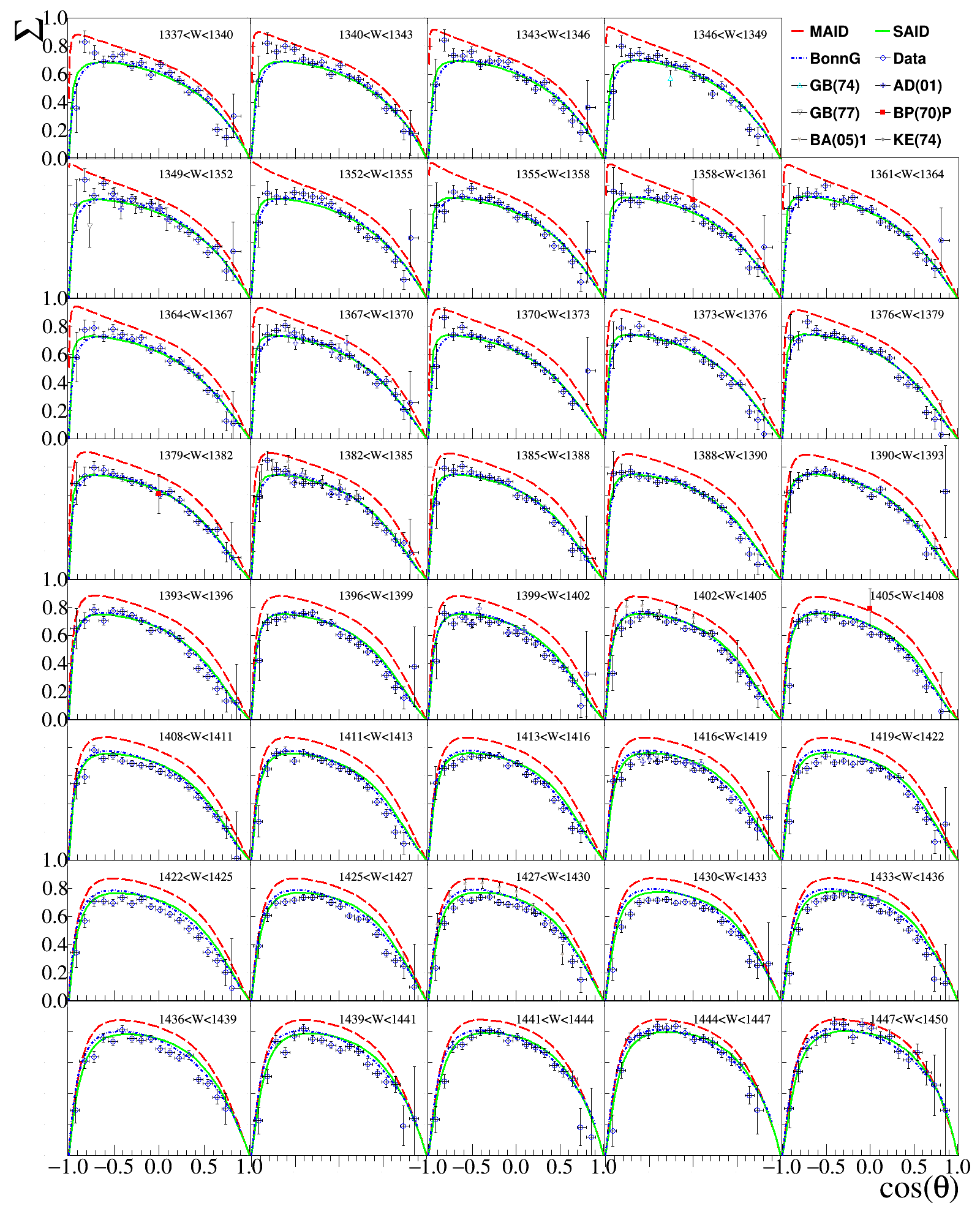}
\caption{(Color online) Photon asymmetry (open blue circles) as a function of $\cos\theta_{cm}$. The $W$ range for each plot is shown on the top right. Predictions from PWAs (MAID \cite{kamalov2008recent}, SAID \cite{DCSA2}, Bonn-Gatchina \cite{Anisovich2010photoproduction}) are shown as colored lines and results from previous experiments (AD(01) \cite{Adamianmeasurement}, BA(05)1 \cite{Bartalini2005measurement}, BP(70)P \cite{zdarko1972asymmetry}, GB(74) \cite{gorbenko1974proton}, GB(77) \cite{gorbenko1977double}, KE(74) \cite{knies1974measurement}) see legend, are taken from the SAID database \cite{saidweb}.}
\label{fig:SigmaW2}
\end{figure*}

The results of the $\mathrm{\Sigma}$ measurements are presented in figs. \ref{fig:SigmaW1} and \ref{fig:SigmaW2} as binned $\cos\theta_{cm}$ distributions across the $W$ range 1.214-1.450 GeV, where the uncertainty in the degree of polarization is small compared with the magnitude of the signal. In total 1403 new measurements of $\mathrm{\Sigma}$ are presented. Each data bin is shown alongside any previous data found on the SAID database \cite{saidweb} that lies within the energy range of the bin. In many cases the W-binning of previous data, which is not shown on the figures, was much larger than the present work. This needs to be borne in mind in comparing the statistical accuracy of the previous data with the current measurements.

It is clear that the new photon asymmetry measurements represent a significant improvement in precision in this energy range and are an important addition to the world data pool. The high statistics also provide the opportunity to carry out a \textit{moment analysis}, where angular distributions of the profile function of the beam asymmetry, $\check{\mathrm{\Sigma}}$ (= $\sigma_{0}\mathrm{\Sigma}$) are fitted with associated Legendre polynomials. Comparison of the fitted coefficients with model predictions can then be used to draw inferences about the partial wave contributions. For a full description of this method, and its application to recent photoproduction data, we refer the reader to \cite{LmaxPaper} (in preparation, to be submitted for publication) with similar work carried out in \cite{Dugger-PhysRevC.88.065203} along with results for center of mass energies $W=1700-2100$MeV. As outlined in these works, the profile function can be expressed as
\begin{equation}
\check{\mathrm{\Sigma}} \left(W,\theta\right) = \frac{q}{k} \sum_{n = 2}^{2 \ell_{\mathrm{max}}} a^{\mathrm{\Sigma}}_{n} \left(W\right) P^{2}_{n} \left(\cos \theta\right) \mathrm{,} \label{eq:SigmaDistrPara}
\end{equation}

where $P_{n}^{m} (\cos \theta)$ are associated Legendre polynomials, $a^{\mathrm{\Sigma}}_{n} \left(W\right)$ are the energy dependent Legendre coefficients and $\frac{q}{k}$ is a Lorentz invariant 2-body phase space factor. The strategy is to increase the truncation order $\ell_{\mathrm{max}}$, starting from $\ell_{\mathrm{max}} = 1$, performing repeated fits using (eq. \ref{eq:SigmaDistrPara}) until the $\chi^{2} / \mathrm{ndf}$ is satisfactory (i.e. as close to 1 as possible) and is not improved further by increasing the truncation limit $\ell_{\mathrm{max}}$. This way, one gets an indication of the dominant partial wave contributions by looking at the angular distributions of the profile function. The procedure is fully model-independent and, furthermore, reliably extracts the $\ell_{\mathrm{max}}$ of the dominant partial waves contributing to an observable \footnote{Interferences between dominant lower partial waves and suppressed higher partial waves $\ell_{\mathrm{max}}\ge$ 3, which may still be important for a full multipole analysis, can still provide contributions to lower order coefficients and, hence, remain hidden from this analysis approach.}.

\begin{figure}
\includegraphics[width=0.49\textwidth]{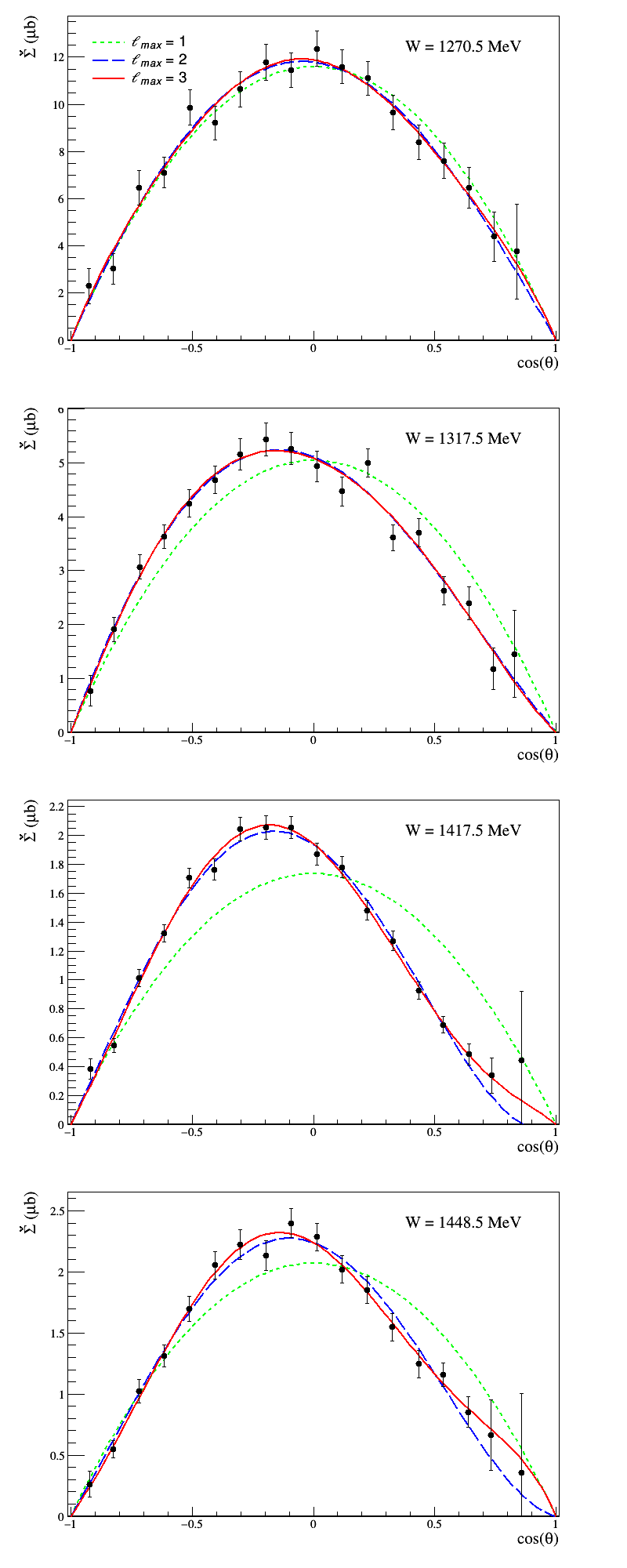}
\caption{(Color online). Examples of truncated Legendre polynomial fits to the angular distributions of the profile function $\check{\mathrm{\Sigma}}$}
\label{fig:profilefits}
\end{figure}

\begin{figure}
\includegraphics[width=0.49\textwidth]{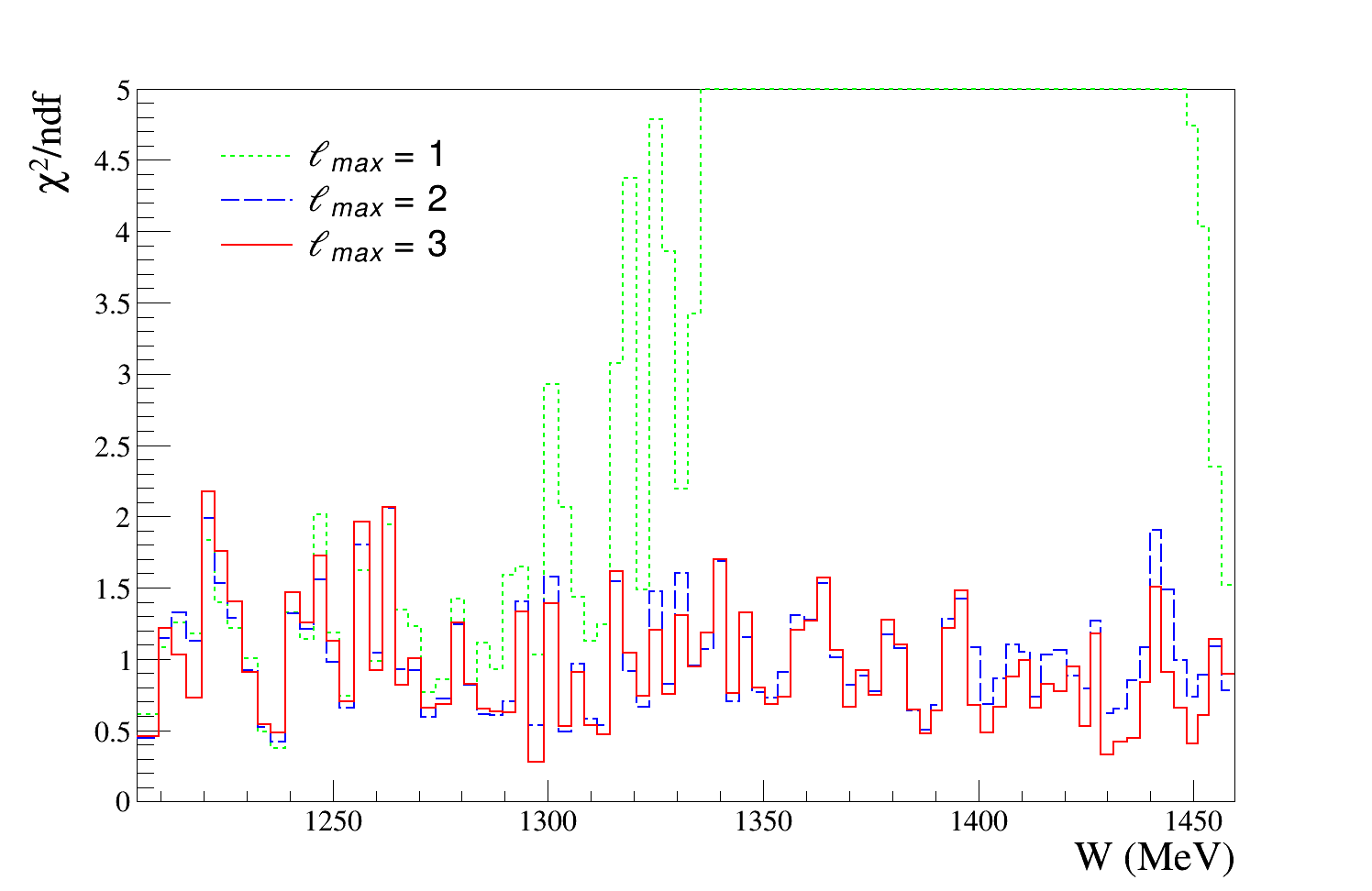}
\caption{(Color online) $\chi^2/ndf$ of Legendre polynomial fits to each $W$ bin for different truncation orders.}
\label{fig:Chi2}
\end{figure}

In order to evaluate the profile function, data for the unpolarized differential cross section $\sigma_{0}$ are needed. For this purpose, we chose the recent $\pi^{0}$-data measured by the A2-collaboration \cite{DCSA2}. For each kinematic bin $(W, \theta)$ in this work, data at nearest neighboring kinematic points were selected from the $\sigma_{0}$ dataset for the evaluation of the profile function. Standard rules for error propagation were applied. Example fits are shown in fig. \ref{fig:profilefits} and the result of the moment analysis is summarized in fig. \ref{fig:Chi2}, showing the resulting $\chi^{2}/\mathrm{ndf}$ for different truncation angular momenta  plotted vs. energy $W$. It is seen that in the low energy region, up to $W \simeq 1300 \hspace*{1.5pt} \mathrm{MeV}$, a truncation at the $P$-waves ($\ell_{\mathrm{max}} = 1$) can already describe the data. However, going beyond $1300 \hspace*{1.5pt} \mathrm{MeV}$ one has to truncate at least at the $D$-waves ($\ell_{\mathrm{max}} = 2$), while the inclusion of $F$-waves ($\ell_{\mathrm{max}} = 3$) can still make a small improvement to the fit in a few bins. On the basis of these fits we conclude that our dataset is dominated by $S$- and $P$-waves in the lower energy region, while the higher region shows significant modifications due to $D$-waves. For further interpretation we consider the fitted Legendre coefficients and compare with calculations from the Bonn Gatchina group \cite{Gutz2014}.

\begin{figure*}
\includegraphics[width=1\textwidth]{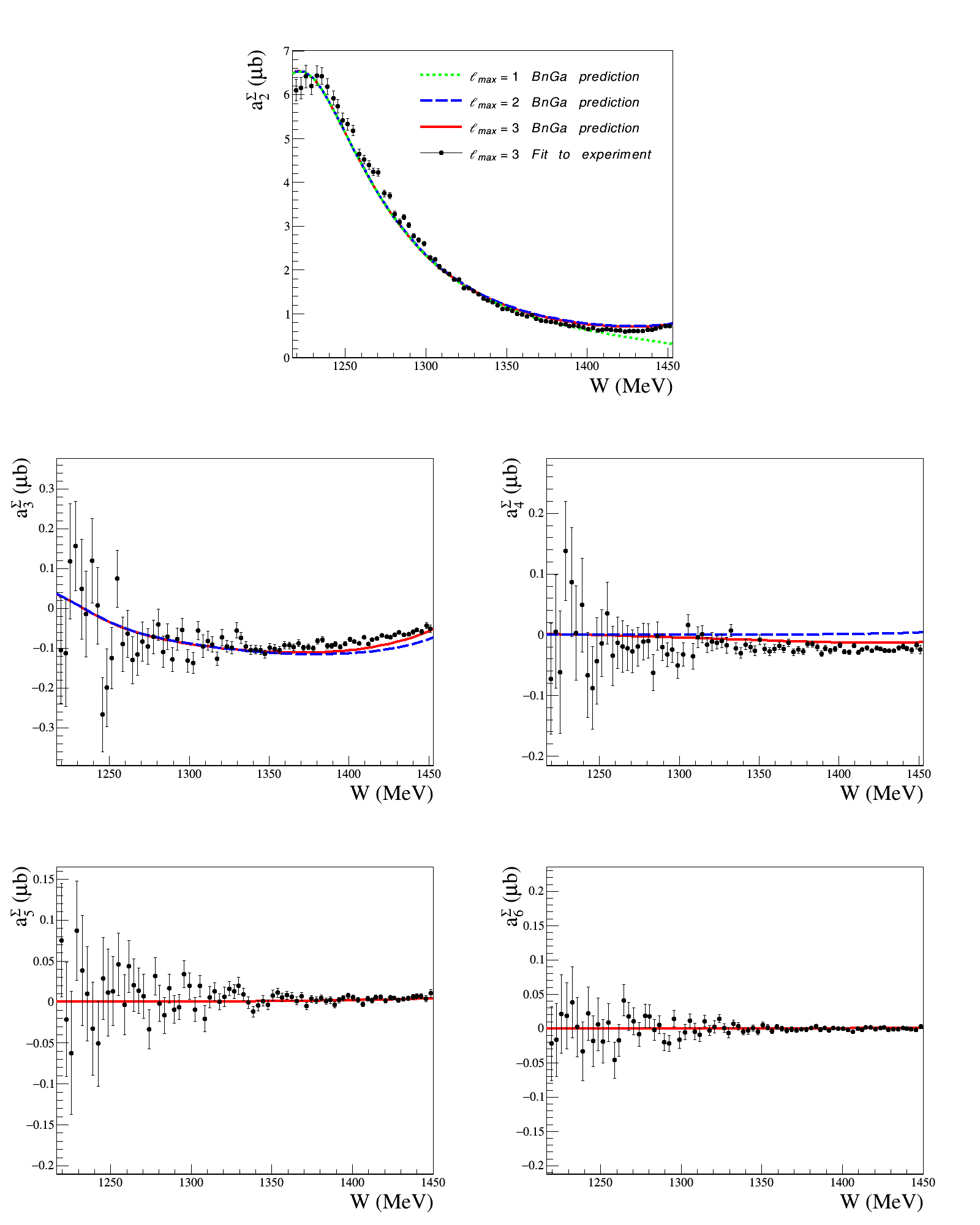}
\caption{(Color online) Legendre coefficients $a^{\mathrm{\Sigma}}_{2,...,6}$ extracted from fits to the profile function $\check{\mathrm{\Sigma}}$. The coefficients shown as filled circles are from an $\ell_{max}=3$ truncated fit, plotted alongside the Bonn-Gatchina predictions (curves) for different $\ell_{\mathrm{max}}$ truncations where the coefficient indices are limited by eq. \ref{eq:SigmaDistrPara}.}
\label{fig:Coef}
\end{figure*}

Figure \ref{fig:Coef} shows the results for the fitted Legendre coefficients. For truncations up to $\ell_{\mathrm{max}} = 3$ the angular distribution (eq. \ref{eq:SigmaDistrPara}) takes the shape
\begin{align}
 \check{\mathrm{\Sigma}} \left(W,\theta\right) =& \frac{q}{k} \big( a_{2}^{\mathrm{\Sigma}} (W) P_{2}^{2} (\cos\theta) + a_{3}^{\mathrm{\Sigma}} (W) P_{3}^{2} (\cos\theta)  \nonumber \\
 & + a_{4}^{\mathrm{\Sigma}} (W) P_{4}^{2} (\cos\theta) + a_{5}^{\mathrm{\Sigma}} (W) P_{5}^{2} (\cos\theta) \nonumber \\
 & + a_{6}^{\mathrm{\Sigma}} (W) P_{6}^{2} (\cos\theta) \big) \mathrm{.} \label{eq:SigmaProfFunctLmax3Decomposition}
\end{align}
Furthermore, the composition of the $a_{n}^{\mathrm{\Sigma}}$ in terms of multipoles can be written in a symbolic notation (described in more detail in \cite{LmaxPaper}) as follows:
{
\allowdisplaybreaks
\begin{align}
 a_{2}^{\mathrm{\Sigma}} &= \left< S,D \right> + \left< P,P \right> + \left< P,F \right> + \left< D,D \right> + \left< F,F \right> \mathrm{,} \label{eq:a2PartialWaveDecomposition} \\
 a_{3}^{\mathrm{\Sigma}} &= \left< S,F \right> + \left< P,D \right> + \left< D,F \right> \mathrm{,} \label{eq:a3PartialWaveDecomposition} \\
 a_{4}^{\mathrm{\Sigma}} &= \left< P,F \right> + \left< D,D \right> + \left< F,F \right> \mathrm{,} \label{eq:a4PartialWaveDecomposition} \\
 a_{5}^{\mathrm{\Sigma}} &= \left< D,F \right> \mathrm{,} \label{eq:a5PartialWaveDecomposition} \\
 a_{6}^{\mathrm{\Sigma}} &= \left< F,F \right> \mathrm{.} \label{eq:a6PartialWaveDecomposition}
\end{align}
}
In this shorthand notation, each scalar product symbol $\left< - , - \right>$ denotes all occurring interference terms among multipoles of definite $\ell$-quantum-numbers. For instance, $\left< S,D \right>$ denotes a sum:
\begin{equation}
\left< S,D \right> = \sum_{\mathcal{M},\mathcal{M}^{\prime}=\left\{E,M\right\}} \sum_{p,p^{\prime} = \left\{\pm\right\}} c_{p,p^{\prime}}^{\mathcal{M},\mathcal{M}^{\prime}} \mathrm{Re}\left[\mathcal{M}_{0 p}^{\ast} \mathcal{M}^{\prime}_{2 p^{\prime}}\right] \mathrm{.} \label{eq:ExampleSum}
\end{equation}
To interpret the distribution plots we evaluated the Legendre coefficients $a^{\mathrm{\Sigma}}_{(2,\ldots,6)}$ using multipoles from the Bonn-Gatchina solution BnGa 2014-02 \cite{Gutz2014}. Different lines in the plots denote the Legendre coefficients, evaluated using BnGa-predictions only up to and including $P$-, $D$- and $F$-waves. Hence, the predictions have also been truncated, in order to study the influence of different partial wave interferences in the model.

Firstly, we note that there is good agreement between the Legendre coefficients with the BnGa-curves. This is encouraging, since the data analyzed in this work have not yet been fitted by the Bonn-Gatchina group. Furthermore, the Legendre coefficients coming in with the $F$-waves, i.e. $a^{\mathrm{\Sigma}}_{5}$ and $a^{\mathrm{\Sigma}}_{6}$, are consistent with zero when looking at the fits extracted from the $\check{\mathrm{\Sigma}}$ data, as well as the model predictions. Therefore everything is consistent with the interpretation that in the energy regime considered here, the $F$ waves themselves are quite small, (observe that in a truncation at $\ell_{\mathrm{max}} = 3$, the coefficient $a_6^{\Sigma}$ is a pure $\left<F,F\right>$-term). However, they are not totally unimportant. This can be seen by looking at $a^{\mathrm{\Sigma}}_{3}$ and $a^{\mathrm{\Sigma}}_{4}$. Both coefficients should be zero (logically) for the model curve up to $P$-waves. Including the $D$-wave multipoles into the evaluation of the model prediction brings the BnGa curves closer to the measured data.

However, a further significant improvement of the description can be reached by including the BnGa $F$-waves. At first glance, this seems surprising since no well established (PDG three-star or higher rated) resonance is known to exist in the energy region of this work. However, inspection of the multipole compositions eq. \ref{eq:a3PartialWaveDecomposition} and eq. \ref{eq:a4PartialWaveDecomposition} shows that the $F$-waves enter both of them via interference terms. For the coefficient $a^{\mathrm{\Sigma}}_{3}$, those are $\left< S,F \right>$- and $\left< D,F \right>$-terms. Well known $S$- and $D$-wave resonances within the reach of our data are of course the $N(1535) \frac{1}{2}^{-}$ and $N(1520) \frac{3}{2}^{-}$. The improvement to the agreement between the model and the measurements when the truncation is extended to $\ell_{\mathrm{max}} = 3$ strongly suggests that there is  interference between these two resonances with the very small $F$-wave contribution. The quantity $a^{\mathrm{\Sigma}}_{4}$ on the other hand has a $\left< P,F \right>$-term. Therefore, it is sensible to assume that some interference with the Roper resonance $N(1440) \frac{1}{2}^{+}$ also comes into play. 

To summarize, the data for the beam asymmetry $\mathrm{\Sigma}$ analyzed in this work show dominant contributions up to $\ell_{\mathrm{max}} = 2$ and are in good agreement with model predictions. A small additional improvement at $\ell_{\mathrm{max}} = 3$ indicates the possibility of interference of the $F$-wave contribution with the $N(1520) \frac{3}{2}^{-}$, $N(1535) \frac{1}{2}^{-}$ and $N(1440) \frac{1}{2}^{+}$ resonances.

\newpage
\clearpage

\newpage
\clearpage

\section{Conclusion}
We have presented new, high statistics, measurements of the photon asymmetry $\mathrm{\Sigma}$ for the $\overrightarrow{\gamma}$p$\rightarrow\pi$$^{0}$p reaction in the range $W$=1214-1450 MeV, taken with the MAMI A2 real photon beam and CrystalBall/TAPS detector systems. The results are compared with MAID, SAID, and Bonn-Gatchina PWA predictions, together with the world dataset. There is good agreement with previous  $\mathrm{\Sigma}$  measurements from Mainz, GRAAL, and Yerevan in regions where there is overlap. This study additionally provides new high statistics measurements in kinematic regions not covered by these previous experiments. We have been able to use this high statistics data together with recently measured cross-sections to carry out a \textit{moment analysis}, fitting  the angular distributions of the profile function of the beam asymmetry, with associated Legendre polynomials. A comparison with calculations from the Bonn-Gatchina model shows that the precision of the data is good enough to further constrain the higher partial waves, and there is an indication of interference between the very small $F$-waves and the $N(1520) \frac{3}{2}^{-}$,  $N(1535) \frac{1}{2}^{-}$ and $N(1440) \frac{1}{2}^{+}$ resonances.

\label{conclusion}

\section*{Acknowledgements}
\label{acknowledgements}
This paper is dedicated to the memory of Bob Owens (University of Glasgow), who passed away in 2015 and had led the development of the tagged photon facility at MAMI.
The authors wish to acknowledge the outstanding support of the accelerator group and operators of MAMI. 
This work was supported by the UK Science and Technology Facilities 
Council (ST/J00175/1, ST/G008604/1, ST/G008582/1, ST/J00006X/1)
Deutsche Forschungsgemeinschaft (SFB 443, SFB/TR 16), 
European Community-Research Infrastructure Activity (FP6), 
Schweizerischer Nationalfonds, the U.S. Department of Energy (Office of Science, Office of
Nuclear Physics, under Award No. DE-SC0014133) and NSF, 
and Canadian NSERC (FRN: SAPPJ-2015-00023).

\bibliographystyle{epj.bst}
\bibliography{thesis}

\newpage
\clearpage

%
%

\end{document}